\documentclass[epj,twocolumn]{webofc}

\usepackage[varg]{txfonts}   % Web of Conferences font
\usepackage{siunitx}
\usepackage{hyperref}
\usepackage[tight,nice]{units}

\usepackage{aas_macros}

\newcommand{\del}[1]{}
\newcommand{\R}{\ensuremath{\mathcal{R}}}
\newcommand{\n}{{\rm n}}
\newcommand{\Hg}{{\rm Hg}}
\newcommand{\diff}[1]{\operatorname{d}\ifthenelse{\equal{#1}{}}{\,}{\!#1}}

\newcommand{\pow}[2]{\ensuremath{#1\!\times\!10^{#2}}}

\newcommand{\ecm}{\ensuremath{\si{\elementarycharge}\!\cdot\!\cm}}
\newcommand{\dn}{\ensuremath{d_\text{n}}}

\newcommand{\cm}{\ensuremath{\mathrm{cm}}}

\newcommand{\magHg}{\texorpdfstring{\ensuremath{{}^{199}\text{Hg}}}{mercury-199}\xspace}

\woctitle{Flavour changing and conserving processes}

\begin{document}

\title{Search for electric dipole moments}
\author{Klaus~Kirch\inst{1,2}\fnsep\thanks{\email{klaus.kirch@psi.ch}} \and
Philipp Schmidt-Wellenburg\inst{1}\fnsep\thanks{\email{philipp.schmidt-wellenburg@psi.ch}} 
}
\institute{Paul Scherrer Institut, CH-5232 Villigen PSI,
Switzerland
\and
ETH Z\"{u}rich, Institute for Particle Physics and Astrophysics, CH-8093 Z\"{u}rich, Switzerland  }
\abstract{Searches for permanent electric dipole moments of fundamental particles and systems with spin are the experiments most sensitive to new CP violating physics and a top priority of a growing international community. We briefly review the current status of the field emphasizing on the charged leptons and lightest baryons.
}
\maketitle

\section{Introduction}
\label{sec:Intro}
\begin{figure}
\centering
\includegraphics[width=0.92\columnwidth]{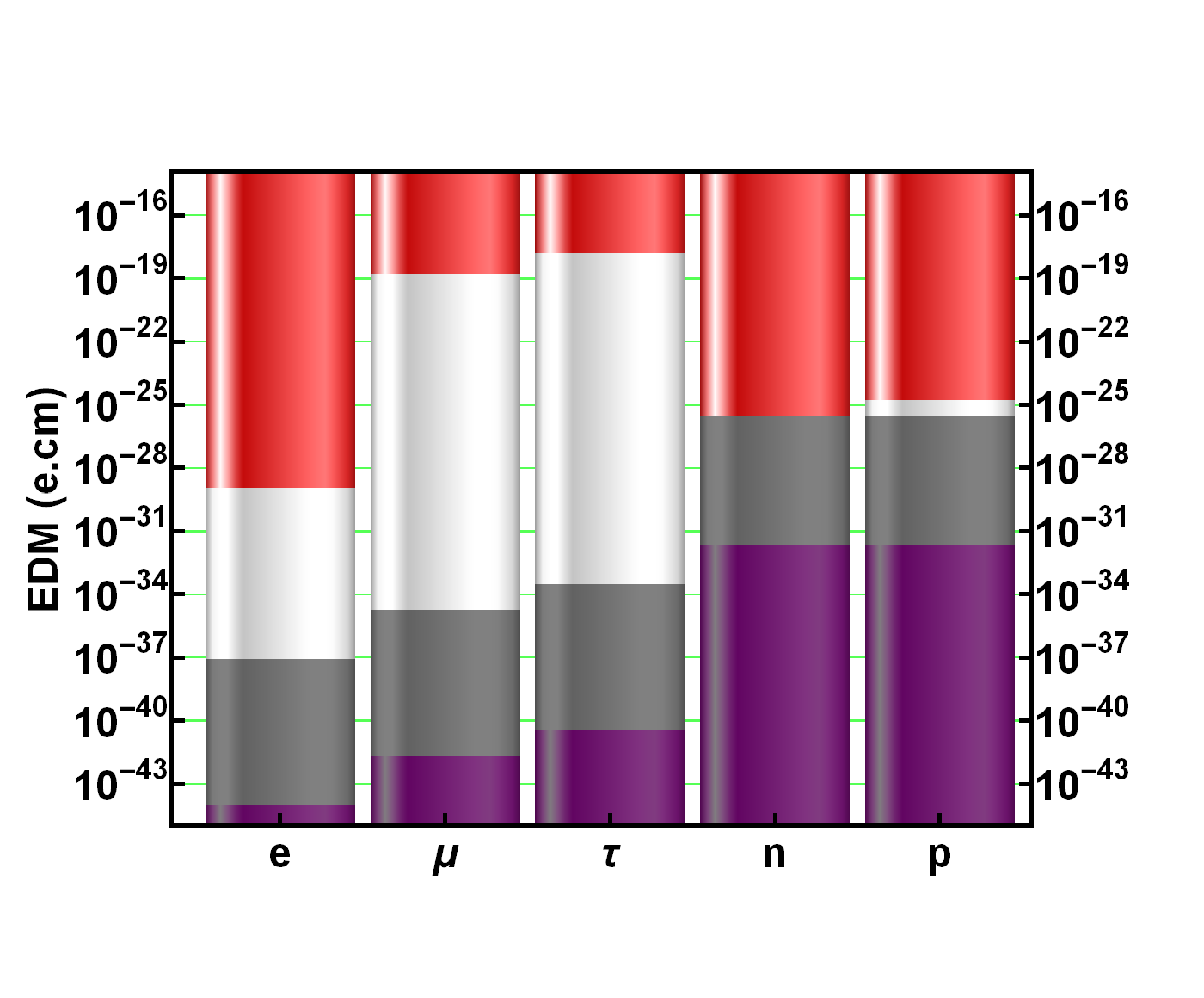}
\vspace{-1cm}
\caption{Current most stringent limits on the EDM (red) of the charged leptons and the lightest baryons displayed along the expected background from the CP-violating CKM-phase of the weak interaction (violet) and  possible strong CP-violation (grey). The strong CP-violation assumes that $\theta$ is as large as allowed by the current limit on the neutron or \magHg{} EDM. Note that the limits on the electron and proton EDM are extracted under the sole source assumption from the measurements on ThO and \magHg{}, respectively, see Tab.~\ref{tab:limits}. The limit on the tau uses that of the electron. The CKM contributions are the estimated fourth order loop contributions to the electron~\cite{Pospelov2014} and scaled for muon and tau. 
The CKM contributions to neutron and proton may vary by an order of magnitude or so, compare e.g.~\cite{Pospelov2005}.
(Figure courtesy of P. Mohanmurthy)}.%
\label{fig:EDMOverview}%
\end{figure}

The discovery of a permanent electric dipole moment of a particle would be an unambiguous signal of the violation of parity (P) and time reversal (T) symmetries\,\cite{Purcell1950, Landau1957, Ramsey1958}. Invoking the CPT theorem\,\cite{Luders1957} it would also indicate the violation of the combined symmetry of charge and parity (CP). 
At the current level of experimental sensitivity this would in turn also manifest new physics; either as the first measurement of CP-violation (CPV) in the strong sector, described by the $\theta$-term\,\cite{tHooft1976} of quantum chromo dynamics (QCD) of the standard model (SM), or as manifestation of CPV which is naturally part of many beyond SM (BSM) theories.
The known CP violating phase in the CKM matrix of the SM weak interaction so far leads to negligible contributions, at least five to six orders of magnitude smaller than current limits, 
cf.\ Fig.\,\ref{fig:EDMOverview}.

\begin{figure}
    \centering
    \includegraphics[width=0.95\columnwidth]{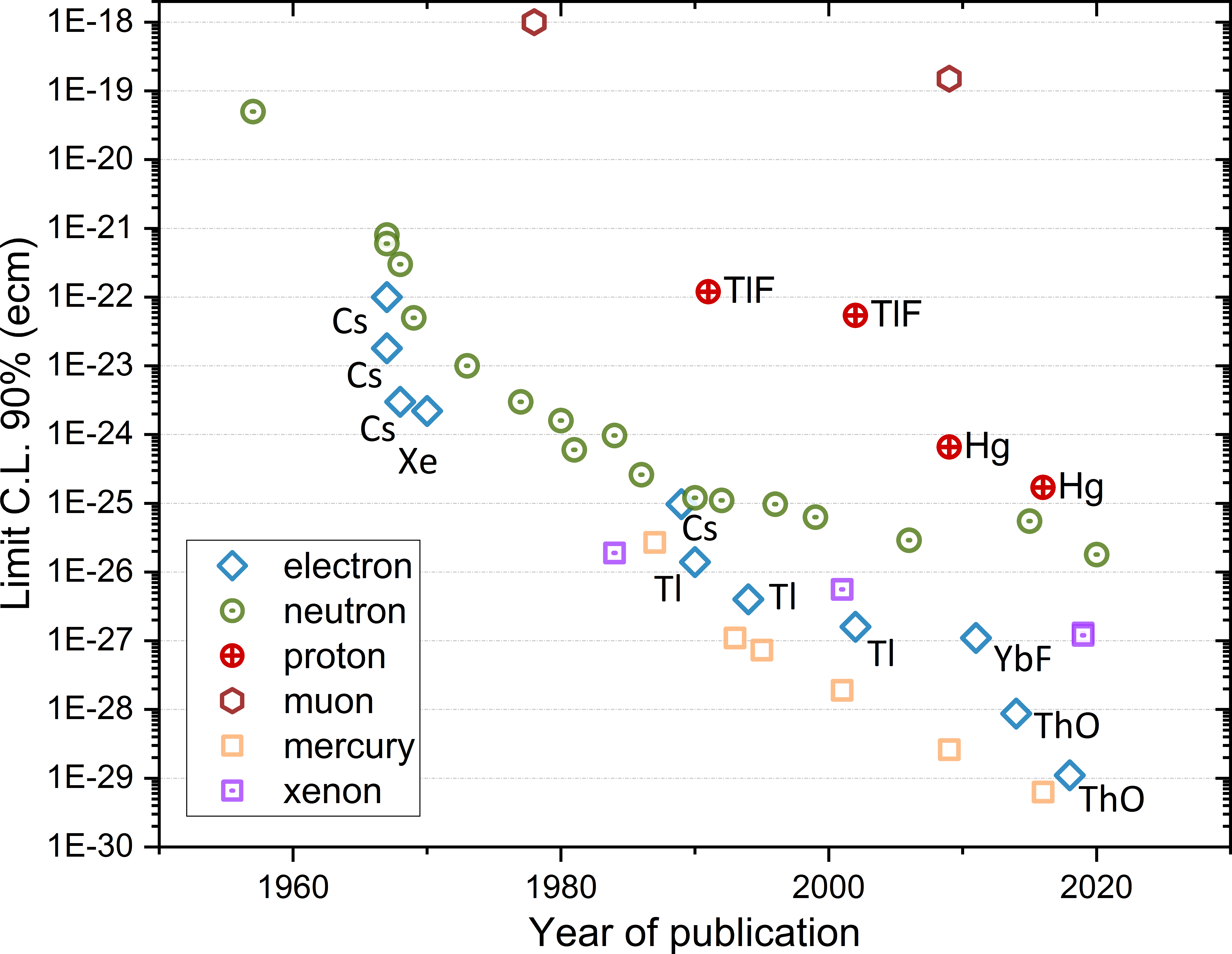}
    \caption{Plot of the history of upper EDM limits (CL 90\%) as function of the year of publication.}
    \label{fig:HistoryPlot}
\end{figure}

Therefore, searches for electric dipole moments constitute a very powerful tool in the quest to find new physics, complementary to e.g. direct searches at the LHC and other future high energy colliders.
Figure\,\ref{fig:HistoryPlot} shows the history of experimental EDM limits and Tab.\,\ref{tab:limits} gives the currently best limits.

The first search for an EDM started as early as 1950, when Smith, Purcell and Ramsey\,\cite{Smith1957} employed the newly developed spin resonance technique with separated oscillating fields\,\cite{Ramsey1950PR} on a beam of neutrons from the reactor in Oak Ridge. The result was only published in 1957 after the discovery of P-violation in $^{60}$Co\,\cite{Wu1957} and in the decay of pions\,\cite{Garwin1957}.
Generically, assuming an unsuppressed CP-violating phase, fundamental fermion EDM today test new physics at mass scales around 10-100\,TeV. A discovery of a finite EDM could help to understand the observed, but unexplained, baryon asymmetry of the Universe\,\cite{Morrissey2012NJP}. The combination of results from different EDM searches, each sensitive to distinct features and aspects of BSM and strong CPV, results in an exceptionally compelling research case for the particle physics community.

\begin{table}
    \centering
    \begin{tabular}{|l|c|c|}
    \hline
         Particle & Method & Upper limit \\
                    &           & (\ecm{} C.L.\,90\%)  \\
         
         \hline
				 electron & ThO$^{\ast}$\,\cite{Andreev2018}& $1.1\times10^{-29}$ \rule{0pt}{2.6ex} \\
         muon & (g-2) storage ring\,\cite{Bennett2009PRD} & $1.5\times10^{-19}$ \\

        tau & eEDM$^{\ast}$\,\cite{Andreev2018,Grozin2009}, updated &  $1.6\times10^{-18}$\\
         neutron & Hg$^\ast$\,\cite{Graner2016PRL} & $1.4\times10^{-26}$ \\
         neutron & UCN storage\,\cite{Abel2020PRL} & $1.8\times10^{-26}$ \\
         proton & Hg$^{\ast}$\,\cite{Graner2016PRL} & $1.7\times10^{-25}$ \\
				 $^{129}$Xe & Xe\,\cite{Sachdeva2019} & $1.2\times10^{-27}$ \\
				 \magHg{} & Hg\,\cite{Graner2016PRL} & $6.3\times10^{-30}$ \\
    \hline    
    \end{tabular}
    \caption{List of most stringent limits on permanent EDMs. Results which are deduced under the ``sole source'' assumption (see text) are marked with an asterisk ($\ast$). }
    \label{tab:limits}
\end{table}

\section{EDM: background free probes of new physics}
\label{sec:Sec2}
Figure\,\ref{fig:EDMOverview} illustrates the power of EDM searches to uncover new physics. The known SM background is limited to the violet fraction coming from the small CP violating phase of the CKM matrix. Hence, any observation of an EDM and in turn new CP-violating physics would be a very significant discovery, irrespective of whether it originates from the QCD sector of the SM or from BSM\@. 
Interpretation of the experimental results requires theoretical treatment on various levels.
Using a set of low energy parameters we can write the EDM of most systems, in particular of atoms or molecules, as a sum\,\cite{Chupp2015PRC},
\begin{align}
    d =&  \sum_{l=e,\mu,\tau}\alpha_{d_l} d_l + \sum_{h=n,p}\alpha_{\bar{d}^{\rm sr}_h}\bar{d}^{\rm sr}_h  + \alpha_{C_{\rm S}} C_{\rm S} + \alpha_{C_{\rm T}} C_{\rm T} \nonumber \\
    &+\alpha_{g_{\pi}^0}\bar{g}_{\pi}^0+\alpha_{g_{\pi}^1}\bar{g}_{\pi}^1
\end{align}
where $\alpha_{d_l}=\partial d/\partial d_l$, and so forth, is the sensitivity of the system to a specific CPV low energy contribution. The low energy constants can be divided into the intrinsic leptonic EDM, $d_{\rm e}$, $d_{\rm \mu}$, and $d_{\rm \tau}$, the T- and P-violating interaction of a scalar or tensor coupling of the electron to the nucleus, $C_{\rm S}$ and $C_{\rm T}$, and the T- and P-violating isoscalar and isotensor pion--nucleon couplings $\bar{g}_{\pi}^0$ and $\bar{g}_{\pi}^1$.

The CPV phase of the weak sector $\delta_{\rm CKM}$ of the SM enters in the case of leptons at fourth loop order and scales with the lepton mass. For the electron, Pospelov and Ritz\,\cite{Pospelov2014} estimated it to be of order $d_{\rm e} = \mathcal{O}(10^{-44})\,e\!\cdot\!\mathrm{cm}$ (replacing old $\mathcal{O}(10^{-38})\,e\!\cdot\!\mathrm{cm}$ estimates by others). In paramagnetic systems the dominant contribution from $\delta_{\rm CKM}$ enters via $C_{\rm S}$ and would result in an electron EDM equivalent value of $\alpha_{C_{\rm S}} C_{\rm S}\sim 10^{-38...-39}\,e\!\cdot\!\mathrm{cm}$, using literature values for $\alpha_{C_{\rm S}}$\,\cite{Chupp2019RMP}. 

The phase $\bar{\theta}$ of the CPV term in QCD enters into these low energy parameters dominantly through the isoscaler pion--nucleon coupling, with $\bar{g}_{\pi}^0 \sim (0.015\pm0.003) \bar{\theta}$\cite{Chupp2015PRC}.

The background by CP violation in the weak sector is essentially irrelevant for today's EDM searches, while the CPV of the strong interaction could result in a signal anytime. For this reason a comprehensive search strategy using many different systems is indicated and necessary for the bigger picture.

For instance, the most sensitive limits on the EDM of the electron come from experiments with molecules like ThO~\cite{Andreev2018} or YbF~\cite{Hudson2011} and molecular ions like HfF$^{+}$\cite{Cairncross2017PRL}. 
These are dominantly sensitive, both, to the intrinsic electron EDM, $d_e$, and to CPV in the scalar electron-nucleon interaction, $C_{\rm S}$. 
Elaborate atomic, molecular and nuclear calculations are required to account for these contributions.
Similarly, when using diamagnetic atoms like  $^{199}$Hg~\cite{Graner2016PRL} or $^{129}$Xe~\cite{Sachdeva2019,Allmendinger2019PRA} nuclear theory is required to extract the CP-violating parameters at the hadronic level, such as the nucleon EDM and CP-violating pion-nucleon couplings, while atomic theory is needed for the extraction of the scalar and tensor couplings of the electron to the nucleons. 

Somewhat easier but still model dependent is
the extraction of the fundamental CP-violating sources (including EDM and chromo-EDM of the quarks) from the EDM measurements of the  neutron, proton or deuteron. One might expect that this will at some point be accessible to precision QCD lattice calcutlations.
Interestingly, as of today, the muon EDM is the only elementary fermion EDM for which the best limit comes from a direct measurement~\cite{Bennett2009PRD}.

The known CPV of the electro-weak part of the standard model produces
EDM only via higher-order loop contributions. These are five orders
of magnitude too small to be detected for current experimental
sensitivities in case of the neutron, eleven orders of magnitude for
the electron and even more for other particles like muons or taus.
However, most new physics scenarios include additional sources of
CP-violation which quite naturally could account for the observed
baryon asymmetry of the universe, and they typically predict much
larger EDM: The experimental EDM bounds thus tightly constrain the
parameter space of such new-physics models and theories.

The muon is of particular interest and is the only fundamental particle which reasonably allows to competitively measure the EDM directly. 
The current best upper limit of the muon EDM, $1.8\times{10}^{-19}e\!\cdot\!\mathrm{cm}$ ($95\%$ C.L.), was obtained parasitically in the ``$(g-2)$'' measurement of the muon at Brookhaven\,\cite{Bennett2006PRD}. This comparatively weak limit leaves the muon EDM as one of the least tested areas of the SM\@.
Recently, Crivellin et~al.~\cite{Crivellin2018} concluded that a muon EDM as large as $3\times{10}^{-22}e\!\cdot\!\mathrm{cm}$
can be obtained in UV complete models possessing an effective decoupling of the muon and electron BSM sectors. Pruna~\cite{Pruna2017} studied leptonic CP violation in the Standard Model Effective Field Theory~(SMEFT). 
Going beyond specific model predictions, he found that several flavor universality violating coefficients connected 
to the second generation are only (and rather weakly) constrained by the muon EDM. 

As hadronic probes, like neutrons and nuclei,
could have EDM induced by the $\theta$-term of QCD, EDM experiments on these systems
can also be considered measurements of $\theta$,
the only remaining parameter of the Standard Model of which only an upper bound exists. 
The fact that hadronic EDM have not been
found so far limits $\theta$ to be extremely, perhaps unnaturally
small (of order $10^{-10}$) which is termed the ``strong CP-problem''.
%
% %%%doesn't seem to fit here%%%%%
%The fact that no fundamental EDM has been found so far also already
%excludes naive supersymmetry models and is known as the ``Susy
%CP-problem''.
%%%%%%

\section{Experimental efforts}
\begin{figure}
\includegraphics[width=\columnwidth]{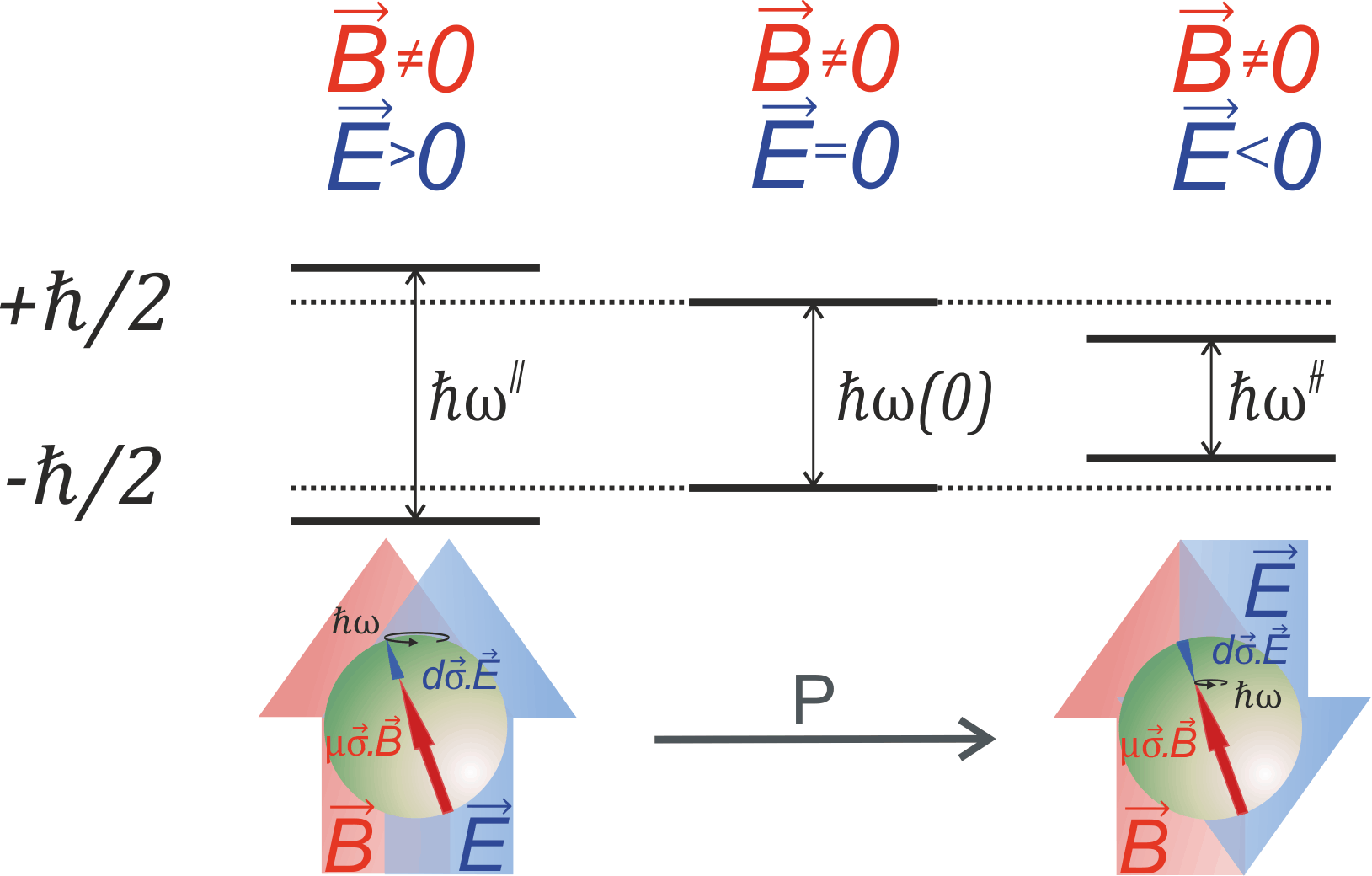}%
\caption{Relative level splitting due to the interaction of the electric dipole moment with an applied electric field. The cartoon illustrates the equivalence of a parity inversion with the inversion of the electric field. Parity inversion changes the Hamiltonian and eigenstates of the system manifesting parity violation.}%
\label{fig:LevelSplitting}%s
\end{figure}

An EDM of a fundamental particle with spin would lead to a level splitting in a strong electric field, see Fig.\,\ref{fig:LevelSplitting}, similar as it is known for the magnetic field. 
This change in the energy level would in turn result in a change of the spin precession frequency.  
By taking the difference between two measurements of the Larmor frequency in configurations where the electric field is parallel ($\omega^{\parallel}$) and anti-parallel ($\omega^{\nparallel}$) to the magnetic field we find that:

\begin{eqnarray}
	\hbar\omega^{\parallel} &= & 2\left|\boldsymbol{\mu_\text{n}\!\cdot\!B}^{\parallel}+\boldsymbol{\dn\!\cdot\!E}^{\parallel}\right| \nonumber\\
	\hbar\omega^{\nparallel} &= & 2\left|\boldsymbol{\mu_\text{n}\!\cdot\!B}^{\nparallel} -\boldsymbol{\dn\!\cdot\!E}^{\nparallel} \right| \nonumber\\
	\dn &= & \frac{\hbar\left(\omega^{\parallel}-\omega^{\nparallel} \right)-2\mu_\text{n}\left(B^{\parallel}-B^{\nparallel}\right)}{2\left(E^{\parallel}-E^{\nparallel}\right)}.
\label{eq:DiffConfig}
\end{eqnarray}
The instantaneous sensitivity of a measurement to an electric dipole moment is
\begin{equation}
		\sigma(d) \propto \frac{1}{PE\sqrt{\dot{N}}T^{3/2}A},
\label{eq:sensitivity}
\end{equation}
where $P$ is the initial polarization, $E$ the electric field strength, $\dot{N}$ the %number of detected final states per unit time, $T$ the measurement time, and $A$ the
number of detected final states per (coherent) measurement time $T$, and $A$ the
analysis power of the final state.
In general the required measurements in equation\,\eqref{eq:DiffConfig} are made in two adjacent volumes with opposite electric fields ($E^{\parallel}=-E^{\nparallel}$) inside the same magnetic field ($B^{\parallel}-B^{\nparallel} = 0$), or by measuring first one configuration then the other and changing the polarity of the electric field from $E^{\parallel}$ to $E^{\nparallel}=- E^{\parallel}$ (or analogously with $B$) inbetween. In the first case it is of paramount importance to make sure that the two spatially separated measurements have the same magnetic field configuration (no or small magnetic-field gradients), while in the second case it is essential to make sure that the magnetic field is stable and/or precisely monitored in time. 
In addition to highest instantaneous sensitivity experimental techniques need to guarantee, that the variation of $\left(B^{\parallel}-B^{\nparallel}\right)$ over $T$ is much smaller $\sigma(d)$.

\subsection{Searches for hadronic EDM}
Searches using baryons and nuclei are particularly sensitive to CPV arising from the QCD $\theta$-term\,\cite{Vries2015PRC}. 

While the current and previous limits on EDM of nuclei and the proton are inferred from measurements of diamagnetic atoms, direct limits exist for the neutron~\cite{Abel2020PRL} and the $\Lambda$ hyperon\,\cite{Pondrom1981}.
Regarding the important plans to measure EDMs for heavy baryons, {\bf we refer to the contribution by Nicola Neri in these proceedings~\cite{}}.

The EDM of the neutron has been searched for in a long series of experiments.
Experiments using neutron beams were abandoned in the 1970s when the systematic effect of a relativistic magnetic field $B_{\rm rel} = \zeta E v/c^2$, where $\zeta$ is the angle between magnetic and electric field, started to dominate the sensitivity\,\cite{Dress1977}. 
Although a new scheme for a competitive neutron beam EDM experiment has been proposed~\cite{Piegsa2013}, most of
today's and next generation experiments use ultracold neutrons\,\cite{Golub1991}. These are neutrons with velocities below about \SI{6}{m/s} which are reflected by suitable material surfaces under all angles of incidence. 
%%%%%%%
%Sufficiently slow neutrons have too low a kinetic energy to overcome the effective optical potential $V_{\rm opt}=2\pi\hbar/m_{\rm n}Nb$, where $N$ is the density and $b$ the bound scattering of the nucleon in the material surface, describing the strong interaction as potential barrier. 
%%%%%%%%%
%%%%%%%%%
This permits the construction of experiments, where neutrons are locked into a macroscopic storage chamber (volume tens of liters) and exposed to strong electric fields $E\approx \SI{1.1}{MV/m}$ for times $T$ up to \SI{180}{s}\,\cite{Abel2020PRL}. Passive magnetic shields made of several layers of an alloy (i.e.\ mu-metal) with a high permeability $\mu$ 
%of about $100000$ 
surround the neutron storage chamber to reduce magnetic field drifts during measurement.
%\,\cite{Fierlinger2012}. 
In the past, remaining magnetic field changes were canceled by taking a relative measurement. Either using a second adjacent precession chamber exposed to the same magnetic field but inverse electric field\,\cite{SerebrovEDM2015} or by using \magHg{} as cohabiting magnetometer\,\cite{Green1998,Pendlebury2015PRD} within the same volume and taking the ratio of neutron to mercury Larmor frequency $ \R=f_\n/f_\Hg $. Future experiments continue to rely on one of these concepts\,\cite{Wurm2019PPNS} or to combine both\,\cite{Abel2018n2EDMProc,Leung2019PPNS}. In addition all spectrometer designs for future measurements include dedicated local magnetic field sensors\,\cite{Abel2019PRAb} to control for systematic effects arising from magnetic field non-uniformities\,\cite{Pignol2012PRA,Abel2019PRA}. 

One requirement for using \magHg{} as cohabiting magnetometer in the neutron EDM search, is that any EDM effect of \magHg is much smaller than the statistical sensitivity of the neutron measurement. This was realized when in 1987 the first limit on the mercury EDM was published\,\cite{Lamoreaux1987} starting a long series of ever more stringent upper limits.
The latest result of the EDM of \magHg{} published in 2016 is $d_{\Hg}=\left(-2.20\pm2.75_{\rm stat}\pm1.48_{\rm sys}\right)\times10^{-30}\ecm$, 
corresponding to an upper limit  of $|d_\Hg|<7.4\times10^{-30}\ecm$ (95\%\,C.L.)\,\cite{Graner2016PRL}. This limit on the \magHg{} atom is the most stringent of all experimental EDM limits.  
\\
The \magHg{} experimnent is  essentially the prototype of all diamagnetic EDM searches. 
It used four cells filled with atomic vapor stacked vertically one upon each other. The two innermost were separated by a plate which was charged to high voltage providing an electric field identical in magnitude but opposite in sign. The outer two cells were kept at ground potential for all times, while the sign of the charged electrode was changed periodically. The atoms were first transversely polarized by optical pumping. Two probe periods, each \SI{20}{s} long, at the beginning and end of a free precession time $T=\SI{170}{s}$ in the dark were used to measure the Larmor frequency. In this way it was possible to reduce otherwise dominant systematic contributions due to noise, depolarisation by probe light, and probe light shifts.

Under the assumption that a possible EDM would be the consequence of only one single CPV source this measurement also yields the most stringent limits on the neutron $|d_{\n\leftarrow\Hg}|<1.6\times10^{-26}\,\ecm$ and proton EDM $|d_{\rm p \leftarrow\Hg}|<2\times10^{-25}\,\ecm$ (all C.L.\ 95\%). As discussed earlier, this is a  pragmatic approach but completely ignores the possibility of the presence of multiple CPV sources for which effects could even cancel. Model independent global analyses as, e.g., discussed in Refs.\,\cite{Chupp2015PRC,Fleig2018JHEP} give a broader view on CPV BSM parameter space going much beyond the ``sole source'' hypothesis. These approaches are consistent with effective field theory analyses of light systems, compare e.g.\,\cite{Dekens2014,Pruna2017}, the extension of which might eventually yield a complete picture.
Nevertheless, while direct measurements of the neutron EDM continue and efforts for the proton and deuteron are being planned, the indirect access to these observables will become better in the future. Before the \magHg{} atom was taking the lead on limiting the proton EDM, the best constraint was coming from a thermal TlF beam experiment~\cite{Cho1991PRA}. New efforts are underway in DeMille's group at Yale to improve the sensitivity to the proton EDM with the CeNTREX experiment using a cold beam and ultimately trapped TlF molecules. 

The latest results from experiments using diamagentic atoms to search for an atomic EDM were reported in 2019 by F.~Allmendiger {\it et al.}\,\cite{Allmendinger2019PRA} $d_{\rm Xe} = (-4.7\pm6.4)\times10^{-28}e$\,cm and N.~Sachdeva {\it et al.}\,\cite{Sachdeva2019} $d_{\rm Xe} = (1.4\pm6.6_{\rm stat}\pm2.0_{\rm sys})\times10^{-28}e$\,cm. These measurements improve the limits on the underlying CPV parameters $g^{0,1}_\pi$ and $\bar{\theta}$ by a factor two and $C_T$ by a factor five.

\begin{figure}
\includegraphics[width=\columnwidth]{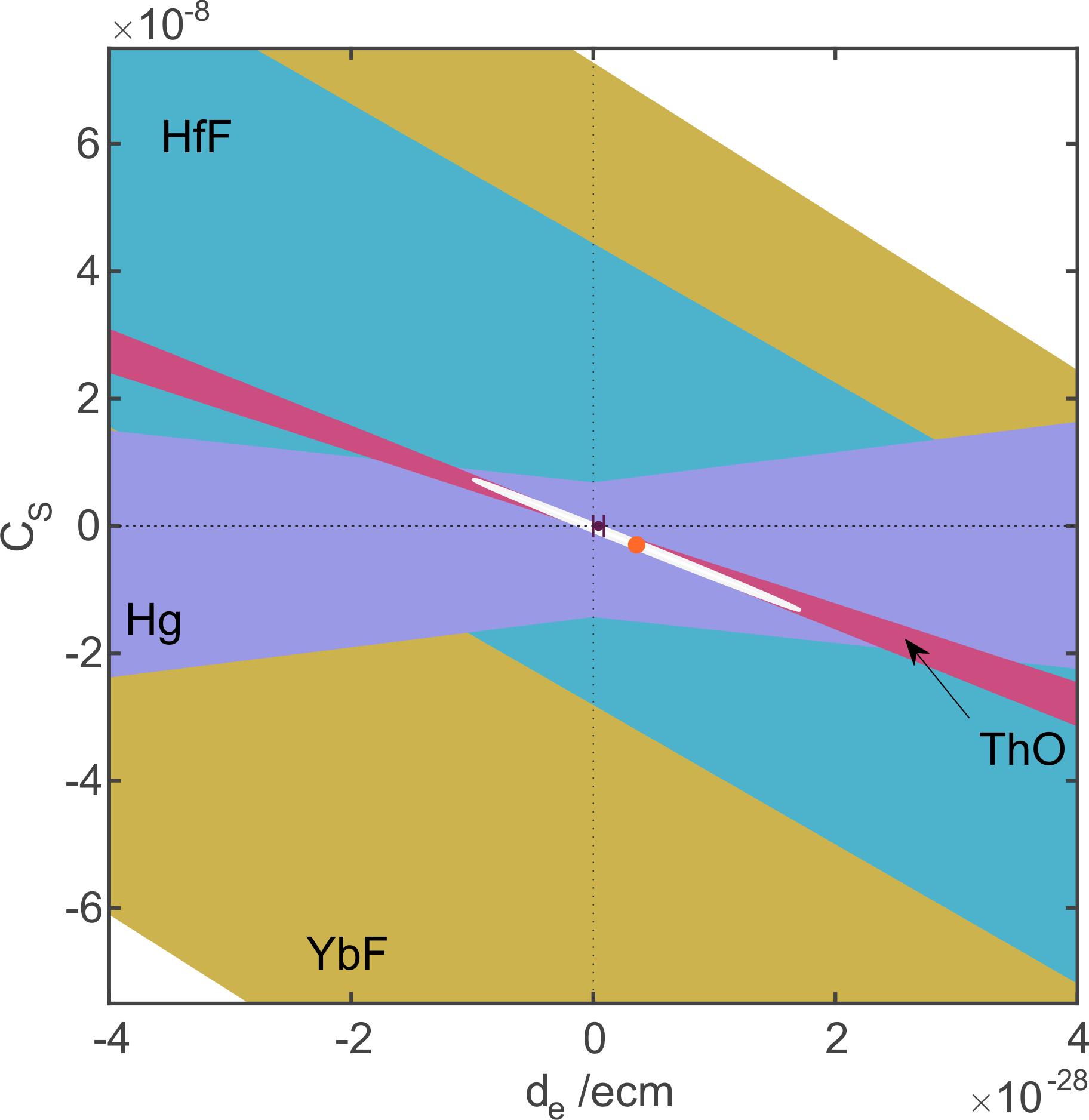}%
\caption{Fit of the current best results from the paramagnetic systems including \magHg{} in order to break degeneracy. The plot is an update of a similar one in~\cite{Fleig2018JHEP}, using the latest results and relations of~\cite{Chupp2019RMP}.
The colored bands denote the results from the different systems, the orange point is the best fit value with a 1-sigma confidence ellipse (white), while the data point with error bars indicates the latest result from the ACME collaboration\,\cite{Andreev2018} on the eEDM assuming only the intrinsic electron EDM as CPV source. }%
\label{fig:EDM-fit}%s
\end{figure}

In section~\ref{sec:Sec2} we have already pointed out that the EDMs of paramagnetic atoms and molecules are mostly sensitive to $d_e$ and $C_S$. However, the correlation of $d_e$ and $C_S$ is similar in paramagnetic systems. It has been pointed out in~\cite{Fleig2018JHEP} that \magHg{} can be used to break the degeneracy. We show an update of that situation in Fig.~\ref{fig:EDM-fit} assuming $d_e$ and $C_S$ as the only sources for the EDMs of these systems. While this is not yet the ultimate analysis, it illustrates the power of using multiple systems.

\subsection{Leptonic EDM}

In passing we mention that EDMs of neutrinos have never been searched for in dedicated experiments. They have been analyzed as side products from other observations and experiments. We refer the reader to the review by Jungmann for more information~\cite{Jungmann2013}.
Also, we are not going to delve deeply into the electron EDM experiments, some of which we mentioned already in section~\ref{sec:Intro}. The leading experimental efforts today use beams of paramagnetic molecules~\cite{Andreev2018, Hudson2011} or stored ions~\cite{Cairncross2017PRL}. The great advantage of polar molecules is in the huge enhancement of the inner-molecular electric field to which the electron is exposed on average. Together with the enhancement factors typical for paramagnetic atoms unique sensitivity to the electron EDM is achieved. In addition to ThO, YbF, HfF, also BaF is being investigated~\cite{Aggarwal2018}. One path forward, which is being actively pursued by almost all experimental collaborations, is to realize cryogenic beams to increase the interaction time and minimize systematic effects, and to eventually optically trap and cool the molecules. Another attempt is to go to heavier systems, involving deformed and even radioactive nuclei, e.g. RaF molecules~\cite{Sasmal2016}. Recently, also the idea to drastically improve the number of systems to be simultaneously probed by applying matrix isolation spectroscopy is being applied to polar molecules~\cite{Vutha2018}.

Instead of the electron EDM efforts, we will rather look closer to the muon EDM.
For convenience, we use a definition of the lepton electric dipole moment in analogy to 
the magnetic dipole moment ($\vec{\mu}=gq\hbar\vec{\sigma}/4mc$) with

\begin{equation} 
		\vec{d_\mu} = \eta\frac{q\hbar}{4mc}\vec{\sigma},
\label{eq:LeptonEDM}
\end{equation}
where $q$, $m$, $\vec{\sigma}$ are the elementary charge, mass, and spin of the lepton, and $\eta$ encodes the interesting physics.

The muon is the only fundamental particle which reasonably allows to measure the EDM directly in a competitive way. The current best upper limit of the muon EDM, see Tab.\,\ref{tab:limits}, was obtained parasitically in the ``$(g-2)$'' measurement in a storage ring at Brookhaven\,\cite{Bennett2006PRD}. 

The spin precession $\vec{\omega}$ of a muon in a storage ring with an electric field $\vec{E}$ and magnetic field $\vec{B}$ is given by:

\begin{equation}
	\vec{\omega}=\frac{q}{m}\left[a\vec{B}-\left(a+\frac{1}{1-\gamma^2}\right)
	\frac{\vec{\beta}\times\vec{E}}{c}\right] +
	\frac{q}{m}\frac{\eta}{2}\left(\vec{\beta}\times\vec{B}+\frac{\vec{E}}{c}\right),
\label{eq:omegaMu1}
\end{equation}

\noindent where $a=(g-2)/2$ is the anomalous magnetic moment~\cite{Bennett2006PRD} of the muon, and $\gamma=1/\sqrt{1-\beta^2}$. The first term of equation\,\eqref{eq:omegaMu1}, is the anomalous precession $\vec{\omega}_{\rm a}$, the difference of the Larmor precession frequency and the cyclotron frequency, oriented parallel to the magnetic field. The second term is the precession $\vec{\omega}_{\rm e}$ due to an EDM coupling to the relativistic electric field of the muon moving in the magnetic field $\vec{B}$, with perpendicular orientation to $\vec{B}$.

A dedicated measurement employing the frozen spin technique\,\cite{Farley2004PRL} can significantly increase the sensitivity compared to results obtained as byproduct of the $(g-2)$ measurements. This requires tuning the electric and magnetic field in equation\,\eqref{eq:omegaMu1} such that the first term cancels:

\begin{equation}
	\left[a\vec{B}-\left(a+\frac{1}{1-\gamma^2}\right)
	\frac{\vec{\beta}\times\vec{E}}{c}\right] = 0 . 	 	
\label{eq:FrozenSpinCondition}
\end{equation}

In this case with $\eta =0$, for vanishing or a negligibly small $\mu$EDM, the spin exactly follows the momentum, and in the rest frame of the muon the spin is ``frozen''.
Originally, this idea was proposed for J-PARC using a dedicated new, pulsed, high-momentum muon beam\,\cite{JPARC2009} in 2009. 
However, it has been realized that several orders of magnitude improvement in sensitivity to the $\mu$EDM can be obtained already using existing low momentum muon beams and one muon at a time only~\cite{Adelmann2010JPG}. The important insight was that for the frozen spin condition and $a\beta^2\gamma^2 \ll 1$, 

\begin{equation}
    E \approx a B \beta \gamma^2,
\end{equation}

thus limiting useful muon momenta and $B$ fields by reasonably achievable laboratory electric fields.
Today, at PSI, a project based on the concept outlined in\,\cite{Adelmann2010JPG} is gaining momentum. The experiment could use highly polarized muons, from pion backward decay in flight, with a momentum of \SI{125}{MeV/c}, 
corresponding to a velocity of $\beta c =0.766c\approx \SI{23}{cm/ns}$ from the $\mu$E1 beamline at PSI and a storage ring made of a very uniform, weakly focusing dipole magnet with a field of $\vec{B}=\SI{1.5}{\tesla}$.
In the case of an EDM ($\eta\neq 0$) the spin will start to precess out of the orbital plane building up a net vertical polarization. A tracking system with sufficient directional resolution around the storage ring will detect the decay positrons. The muon  decay asymmetry $\alpha$, will lead to a build-up of an up-down asymmetry with time, as the polarization moves up or down, proportional to $\eta$, the EDM signal.
 
The experimental sensitivity is obtained by modifying equation (5) of Ref.\,\cite{Adelmann2010JPG} by inserting equation\,\eqref{eq:LeptonEDM}:

\begin{equation}
	\sigma(d_{\mu}) =\frac{ \hbar\gamma a}{2P E  \sqrt{N}\tau\alpha},
\label{eq:muEDMsensitivity}
\end{equation}
for a polarization $P$, the muon life time $\tau=\SI{2.2}{\micro\second}$ and the number of detected positrons $N$. The PSI concept requires a radial electric field of $E=\SI{1}{MV/m}$, resulting in a storage ring radius of $r=\SI{0.28}{m}$. 
With an average analyzing power of $\bar{\alpha}=0.3$, an initial polarization of $P=0.9$, and assuming $N=T/(\gamma\tau)=\pow{4}{14}$ muons per year, where $T$ is the effective measurement time in one calendar year, a statistical sensitivity of 
$	\sigma(d_{\mu}) = \unit[\pow{5}{-23}]{\ecm}$ is within reach.

We conclude this contribution by an update of the limit of the EDM of the tau lepton.
Grozin, Khriplovich and Rudenko have shown in 2009~\cite{Grozin2009} how loop contributions of heavier leptons generate an electron EDM. In turn, the limit on the EDM of the electron implies therefore limits for the EDMs of muon and tau. Reference~\cite{Grozin2009} used the 2009 eEDM limit to constrain the EDM of the tau.
The same calculation results in a model independent indirect limit on the muon EDM of 
$|d_\mu|<0.9\times10^{-19}e $\,cm~\cite{Crivellin2018} which is slightly better than the current direct experimental limit. However, this should be considered a rough number only as there could be further contributions and in principle even cancellations. Therefore, we have used the direct limit in Tab.~\ref{tab:limits} to be on the safe side.
The situation is different for the tau for which the update 
with the latest
electron EDM limit~\cite{Andreev2018} yields
\begin{equation}
   |\, d_\tau|<1.6\times10^{-18}e{\rm cm}.
\end{equation} 
This is a more stringent limit on the EDM of the tau by more than an order of magnitude
compared to the one extracted from electron-positron collision data
$e+e^-\longrightarrow\tau^+\tau^-$ for which we calculate
$d_{\tau}<3.4\times10^{-17}e{\rm cm}$ (90\% C.L.) following\,\cite{Matteo2018} in using the real part of the result presented in Ref.\,\cite{Inami2003}.

%\bibliographystyle{woc}
%\bibliography{nEDM-references,UCN-references,BSM-references,SM-references,MuonReferences}

%\input{nEDMLetterReferences}

\end{document}